# Parametric Strong Mode-Coupling in Carbon Nanotube Mechanical Resonators

Shu-Xiao Li,[1,2]† Dong Zhu,[1,2]† Xin-He Wang,[3,4]† Jiang-Tao Wang,[3,4] Guang-Wei Deng,[1,2]* Hai-Ou Li,[1,2] Gang Cao,[1,2] Ming Xiao,[1,2] Guang-Can Guo,[1,2] Kai-Li Jiang,[3,4] Xing-Can Dai,[3,4] Guo-Ping Guo[1,2]*

[1]Key Laboratory of Quantum Information, University of Science and Technology of China, Chinese Academy of Sciences, Hefei 230026, China.

[2]Synergetic Innovation Center of Quantum Information and Quantum Physics, University of Science and Technology of China, Hefei, Anhui 230026, China.

[3]State Key Laboratory of Low-Dimensional Quantum Physics, Department of Physics and Tsinghua-Foxconn Nanotechnology Research Center, Tsinghua University, Beijing 100084, China.

[4]Collaborative Innovation Center of Quantum Matter, Beijing 100084, China.

†These authors contributed equally to this work.

*Corresponding to:gpguo@ustc.edu.cn or gwdeng@mail.ustc.edu.cn

**Abstract:** Carbon nanotubes (CNTs) have attracted much interest for use in nanomechanical devices because of their exceptional properties, such as large resonant frequencies, low mass, and high quality factors. Here, we report the first experimental realization of parametric strong coupling between two mechanical modes on a single CNT nanomechanical resonator, by applying an extra microwave pump. This parametric pump method can be used to couple mechanical modes with arbitrary frequency differences. The properties of the mechanical resonator are detected by single-electron tunneling at low temperature, which is found to be strongly coupled to both modes. The coupling strength between the two modes can be tuned by the pump power, setting the coupling regime from weak to strong. This tunability may be useful in further phonon manipulations in carbon nanotubes.

Outstanding properties have been reported for carbon nanotube mechanical resonators in recent years,[1-16] among which, remarkably, ultrahigh quality factors, as large as $5 \times 10^6$, have been reached.[15] Carbon nanotube resonators are also well-known for their ultralow masses and large and tunable resonance frequencies.[1] Taking advantage of these properties, CNT resonators find wide application in many fields, such as mass and force sensing,[10] magnetic sensing,[9] and parametric amplifications.[5]

A previous study reported strong coupling between the mechanical modes in a single carbon nanotube resonator,[6] in a very specific situation (with suitable gate voltages when the first resonance frequency matches the fraction of the higher order resonances). In addition, on one hand, a pioneering experiment reported coherent phonon manipulation between two modes in a single silicon nitride mechanical resonator.[17] On the other hand, optical tuning[18] and piezoelectric effect tuning[19] of parametric mode coupling in GaAs micromechanical resonators and optical tuning strong modes coupling in a $MoS_2$ resonator[20] have been realized. Moreover, parametric coupling between the first and the second mode in a micromechanical resonator was reported.[21] The question is then raised if there are similar tuning effects in carbon nanotubes.

In this work, we report the first parametric coupling between the two mechanical modes in a single nanotube resonator by applying an extra microwave pump, at arbitrary frequency differences. The mechanical modes are measured by single electron tunneling (SET) at low temperature,[2,3] and both modes are found to be strongly coupled to the SET. By applying extra microwave power, the parametric pump is used to tune the coupling between modes,[17,19,20] where the microwave modulation promotes intra-mode energy exchange. The device structure and the measurement instrument are quite simple, and the coupling strength can easily be tuned from weak to strong by changing the pump power.

Figure 1 shows the sample structure, where a CNT is suspended over two metal (Ti/Au) electrodes [source (*S*) and drain (*D*)]. The trench is designed to be 1.2 μm wide and 200 nm deep, and a metal gate (*g*) 100 nm wide and 50 nm high is located underneath the CNT. Here, the CNT is grown by a chemical vapor deposition method, typically single- or double-walled, 2−3 nm in diameter, and is transferred by a novel near-field micro-manipulation technique, where the CNT can be precisely positioned at the predetermined location.[16] The measurements were performed in a dry refrigerator Triton system with a base temperature of approximately 10 mK and at typical pressures below $10^{-6}$ torr.[22,23] The CNT was biased from the *S* electrode and the DC transport properties were simply measured at the *D* electrode by a commercial multimeter. A bias-T was used to apply both DC and AC voltages on the gate, where the DC voltage was used to tune the fermi level and AC voltage was used to actuate the CNT. On the AC port, a power divider was used to compound two microwave drives ($\omega_p$ and $\omega_d$). There should be many mechanical modes for the nanotube mechanical resonator, in this experiment we choose two distinct modes shown in Fig. 2 (c, d).[6]

Figure 2(a) shows the electron transport properties of the CNT. The CNT behaves as a quantum dot (QD) at large positive gate voltages, while it comes into the Fabry-Perot region for gate voltages smaller than 1 V. Between the two regions there is a ~0.3 V bandgap. We measured the Coulomb diamond of the CNT in the quantum dot region (Fig. 2(b)), obtaining a charging energy of approximately 40-60 meV in the few-electron regime and a gate lever arm of approximately 50%. The large lever arm indicates a high ratio between thegate-QD capacitance and the total QD capacitance.

Driving the CNT with a microwave power $P_d = -30$ dBm, we find two resonance modes ($\omega_1$ for the lower frequency mode and $\omega_2$ for the higher one), measured by single-electron tunneling (see Fig. 2(c)).[2,3] The SET causes a ~$2\pi \times 1.2$ MHz frequency shift for the $\omega_2$ mode and one of ~$2\pi \times 1.5$ MHz for $\omega_1$. The quality factors for both modes are ~50000, and the corresponding decoherence rate $\gamma \sim 2\pi \times 2$ kHz (see the Supplementary Information), indicating a strong coupling between the mechanical motion and the SET.[2,3] Figure 2(d) shows the power dependent response of the two modes, where both modes show Duffing nonlinearity at large power values.[4, 8, 16]

To couple the two modes, we apply an extra pump microwave $\omega_p$ combined with the previous driving power (see Fig. 1). The effect of this pump microwave is to excite both modes and transfer phonons between the two modes (see Fig. 3). Still measured by the SET, the first-order and second-order excitations of both mechanical modes are illustrated in Fig. 3(a), where pump power $P_p = -25$ dBm is applied and the driving power is fixed at $P_d = -40$ dBm. Figure 3(a) shows the DC current as a function of the driving and pump frequencies, where a series of sideband lines are detected in addition to the two base resonance modes $\omega_{1,2}$. These sideband lines indicate vibrations excited by the parametric pump, creating phonons with energies of $\hbar\omega_{1,2} \pm n\omega_p$, where $n$ denotes the $n$th-order excitation. These vibrations indicate the frequency modulation changes the tension of the resonator thus the pump microwave can be treated as a parametric drive. In the following, we only take the first order into consideration, where the Stokes sideband of mode 2 matches the resonance frequency of mode 1, that is, $\hbar\omega_1 \sim \hbar\omega_2 - \hbar\omega_p$ (see black dashed circle in Fig. 3(a)), and the anti-Stokes sideband of mode 1 matches the resonance frequency of mode 2 ($\hbar\omega_2 \sim \hbar\omega_1 + \hbar\omega_p$, see blue dashed circle in Fig. 3(a)). We have illustrated the parametric coupling mechanism in Fig. 3(b), taking the anti-Stokes process as an example.

At higher pump powers (larger than $-20$ dBm), we are able to observe an obvious normal mode splitting between the two modes, where the coupling strength is sufficiently strong (see Fig. 4(a), with a $-10$ dBm pump power). Figure 4(b) and 4(c) show weak coupling (Fig. 4(b)) and strong coupling behavior for the system, with $-25$ dBm and $-5$ dBm pump powers, respectively. Figure 4(d) shows the frequency splitting $\Delta\omega$ for various pump powers, where $\Delta\omega$ can be tuned to as large as $2\pi \times 120$ kHz, indicating a coupling strength $g = \Delta\omega/2$ as large as $2\pi \times 60$ kHz (see Fig. 4(e)). Previously we have estimated the decoherence rates of the two modes to be $\gamma_1, \gamma_2 \sim 2\pi \times 2$ kHz from the fitted quality factors, meaning that the coupling strength is approximately 30 times larger than the resonator line width, indicating a strong coupling regime. Here in our device, the coupling strength as large as $2\pi \times 60$ kHz is about 3 times larger than the silicon nitride devices,[17] two orders larger than the GaAs devices,[19] and of the same order with MoS$_2$ resonators.[20]

Induced by the motion-induced tension, the intrinsic coupling mechanism reported here is the same with previous reported work by Eichler *et al.*[6] However, the coupling method we used here is quite different from the nonlinear coupling method reported before. In that experiment,[6] coupling was caused by a large driving force at frequency close to resonance. When the vibration amplitude of the resonator is in the large regime, nonlinear terms that are the essence of inter-mode coupling cannot be omitted. The benefit of the previous coupling method is that they offered a platform to study interesting nonlinear physics, however, the coupling can only occurred at some special gate voltages there. Compared to their method, our setup is more flexible. Moreover, the coupling strength in our device is highly tunable and the setup is quite simple.

In our device, the resonator was working in the linear regime, and the coupling method between the two vibration modes [we assume the two distinct modes as $X(t)$ and $Y(t)$] is similar to that reported by Okamoto[19] and Chang-Hua Liu *et al.*.[20] We model the frequency response using a standard harmonic oscillator model with a time-varying spring constant. The equations of motion are given by:

$$\ddot{X} + \gamma \dot{X} + \left(\Omega_1^2 + A_1 V_p^2 cos(\omega_p t)\right) X = c(Y - X) + F cos(\omega_d t + \phi), \quad (1)$$

$$\ddot{Y} + \gamma \dot{Y} + \left(\Omega_2^2 + A_2 V_p^2 cos(\omega_p t)\right) Y = c(X - Y) + F cos(\omega_d t + \phi), \quad (2)$$

where γ, $\Omega_{1,2}$, and $c$ represent the damping coefficient, resonance frequency and weak coupling constant between the two distinct vibration modes, respectively. Here $A_{1(2)} V_p^2 cos(\omega_p t)$ is related to the sidebands observed in Fig. 3(a), these sidebands are induced by the pump microwave changing the tension. $c|X - Y|$ is related to the coupling of the two modes, $c$ is an intrinsic constant.

In the experiment, we used a pumping signal with frequency $\omega_p$ being around 1.65 MHz, which is far from the resonance frequencies and large power, approximately 0 dBm, while the driving signal was set to 118 MHz and very low power (−40 dBm) to match the resonance frequency of the lower mode and drive it.

Nonlinear effects were avoided for the low driving power and small mode displacement, while the strong pumping signal $V_p$ with very small frequency $\omega_p$, added to the DC gate voltage $V_g^{DC}$, altered the tension of the carbon nanotube to generate a time-varying spring coefficient.

While the resonance frequency of two orthogonal vibration modes is quite close to each other, we can use the approximation relation $\Omega_2^2 \approx \Omega_1^2 + 2\Omega_1\Delta\Omega$, where $\Delta\Omega$ is the frequency difference between the two modes.

By applying the above approximation, equations (1) and (2) can be rearranged in a matrix form as:

$$\left(\frac{d^2}{dt^2} + \gamma\frac{d}{dt} + \Omega_1^2 + c + \Omega_1\Delta\Omega\right)\begin{pmatrix}X\\Y\end{pmatrix} - \begin{pmatrix}\Omega_1\Delta\Omega & c\\ c & -\Omega_1\Delta\Omega\end{pmatrix}\begin{pmatrix}X\\Y\end{pmatrix} +$$
$$\begin{pmatrix}A_1 V_p^2 \cos(\omega_p t) & 0\\ 0 & A_2 V_p^2 \cos(\omega_p t)\end{pmatrix}\begin{pmatrix}X\\Y\end{pmatrix} = F\cos(\omega_d t + \phi). \qquad (3)$$

The above equation can be diagnosed using the orthogonal matrix U:

$$U = \frac{1}{\sqrt{2\lambda}}\begin{pmatrix}\sqrt{\lambda + \xi} & \sqrt{\lambda - \xi}\\ \sqrt{\lambda - \xi} & -\sqrt{\lambda + \xi}\end{pmatrix}, \qquad (4)$$

where $\xi = \Omega_1\Delta\Omega$ and $\lambda = \sqrt{c^2 + \Omega_1^2\Delta\Omega^2}$.

The above equations can be written using the new basis $\begin{pmatrix}x\\y\end{pmatrix} = U\begin{pmatrix}X\\Y\end{pmatrix}$ as:

$$\ddot{x} + \gamma\dot{x} + \omega_1^2 x + (\Gamma_1 x + \Lambda y)V_p^2 \cos(\omega_p t) = F_1 \cos(\omega_d t + \phi) \qquad (5)$$

$$\ddot{y} + \gamma\dot{y} + \omega_2^2 y + (\Gamma_2 y + \Lambda x)V_p^2 \cos(\omega_p t) = F_2 \cos(\omega_d t + \phi) \qquad (6)$$

where $\omega_{1,2}^2 = \Omega_1^2 + c + \Omega_1\Delta\Omega \mp \sqrt{c^2 + \Omega_1^2\Delta\Omega^2}$, $\Gamma_{1,2} = \frac{A_1+A_2}{2} \pm \frac{(A_1-A_2)\xi}{2\lambda}$, $\Lambda = \frac{(A_1-A_2)c}{2\lambda}$ and $\begin{pmatrix}F_1\\F_2\end{pmatrix} = UF$. Here $\omega_{1(2)}$ directly describe the measured resonant frequencies of the two modes with dynamic coupling, noted as the two measured coupling modes in Fig. 4(a). $\Lambda$ in the equations is related to the coupling.

In our experiment (see Fig. 2(d)), the two coupling mode frequencies $\omega_{1,2}$ are approximately 118 MHz and 119.6 MHz, respectively. The driving microwave frequency was set to match the lower mechanical mode $\omega_1$, and we swept the pumping microwave frequency near the frequency difference between the two coupling modes. A distinct dynamical normal mode splitting was observed when the pumping frequency matched the frequency difference $\Delta\omega = \omega_2 - \omega_1$ (see Fig. 4(c) and (d)). This parametric mode-coupling rate[19] $g$ is given by $g \sim \frac{\Lambda V_p^2}{2\sqrt{\omega_1 \omega_2}}$ from former equations, which means the coupling strength $g$ is proportional to $V_p^2$ (or the pump power in W unit). Shown in Fig. 4(e), the mode coupling rate $g$ is well fitted as a linear function with the pump power in mW unit, verifying that the pump power is a parametric driving and changes the intrinsic resonator tensions, and $g \sim \frac{\Lambda V_p^2}{2\sqrt{\omega_1 \omega_2}}$ from the theory is in good agreement with the experiment data.

We have repeated our experiment in several devices, finding that this kind of parametric strong coupling can be realized in any CNT sample, with arbitrary frequency differences between the two mechanical modes (see the Supplementary Information for another sample).

In conclusion, we have demonstrated parametric coupling between two mechanical modes in a carbon nanotube resonator, measured by strongly coupled single electron tunneling. The method can be used to couple mechanical modes with arbitrary frequency differences. Moreover, we find that the system can be tuned to the strong coupling regime by increasing the pump microwave power. Our research enriches the electron-phonon coupling field and may be useful in further phonon manipulation experiments. Also, our system could possibly be cooled to the quantum regime in the future.[24]

**Acknowledgements**

This work was supported by the National Fundamental Research Programme (Grant No. 2011CBA00200), the National Natural Science Foundation (Grant Nos. 11222438, 11174267, 61306150, 11304301, and 91421303), and the Chinese Academy of Sciences. This work was also supported by the National Basic Research Program of China (2012CB932301), the National Key Basic Research Program of China (MOST 2013CB922003) and the NSF of China (No. 11474178).

Supplementary Information Available:

Further information is provided on the fit of the quality factor and similar results in more devices.

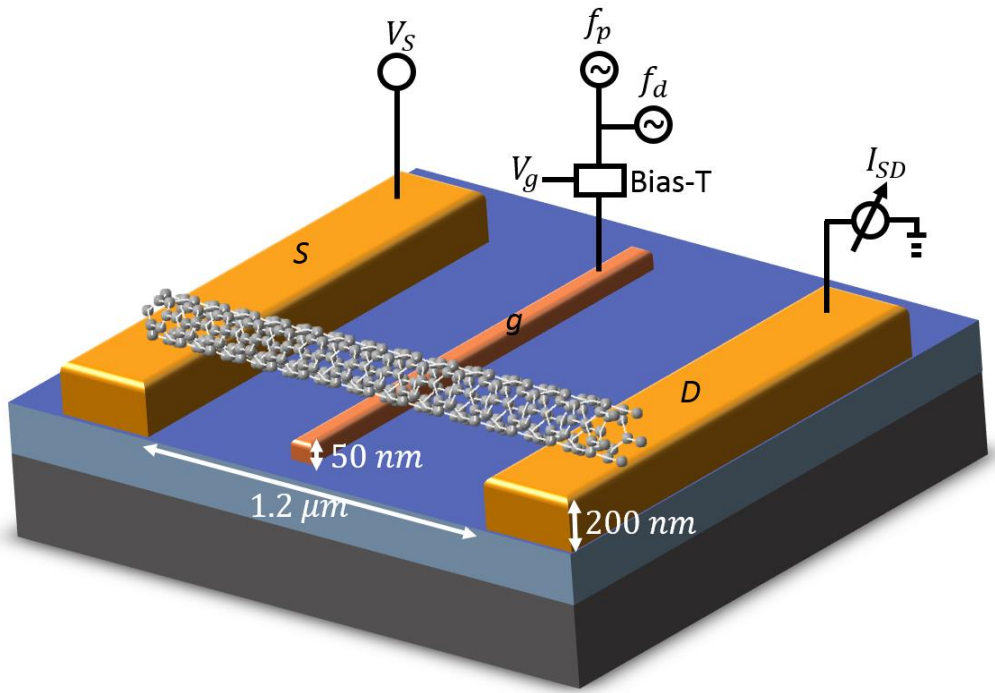

**Figure 1.** Sample structure and measurement schematic. A single- or double-walled carbon nanotube (CNT) is transferred onto two 200-nm-high Ti/Au electrodes, working as source/drain. A 50-nm-high Ti/Au gate under the CNT is used as back gate and can offer DC and AC voltages via a bias-T.

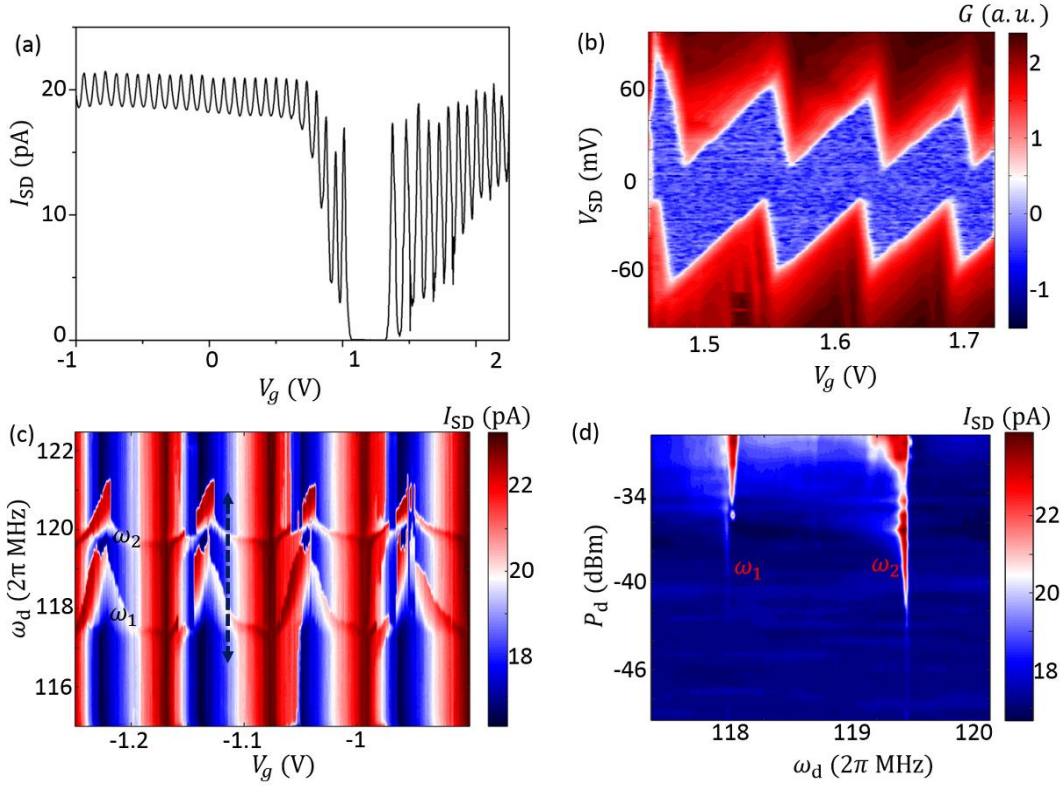

**Figure 2.** Characterization of the basic properties of the device. (a) DC property as a function of gate voltage. Bias between source and drain contacts is 5 mV. (b) Diamond on the electron regime, from which we the charging energy is calculated to be approximately 40 to 60 meV. (c) Measuring the two mechanical modes using single-electron tunneling, where both modes $\omega_1$ and $\omega_1$ show springs near the electron tunneling region. (d) Response of both modes as a function of the driving microwave power. Here the gate voltage is fixed at $-1.12$ V (along the black dashed arrow in (c)).

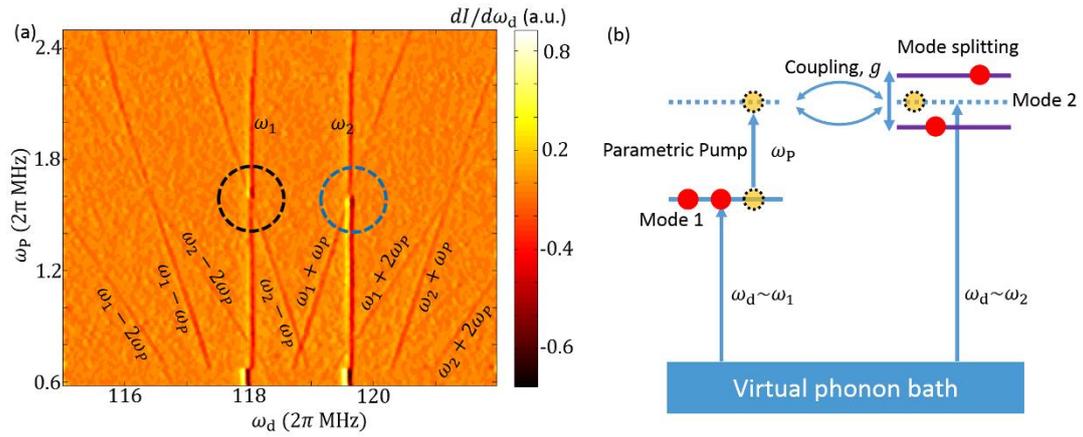

**Figure 3.** Parametric pumping properties of the two modes. (a) Vibrational spectrum of the two fundamental modes as a function of the driving and pump frequencies, with a driving power of $-40$ dBm and a pump power of $-25$ dBm. First- and second-order excitations of both modes can be detected because the modulation pump changes the tension of the resonator. The first-order excitations cross the other mode at $\omega_p \sim 2\pi \times 1.65$ MHz, denoted as the two dashed circles. (b) Schematic of the coupling mechanism for the blue dashed circle in (a). The CNT is seen as a virtual phonon bath and both resonance modes $\omega_1$ and $\omega_2$ are actuated by the driving microwave $\omega_d$. An anti-Stokes process produces phonons from mode 1 with an energy level of $\hbar(\omega_1 + \omega_p)$, thus coupled to mode 2 when $\hbar\omega_2 \sim \hbar\omega_1 + \hbar\omega_p$.

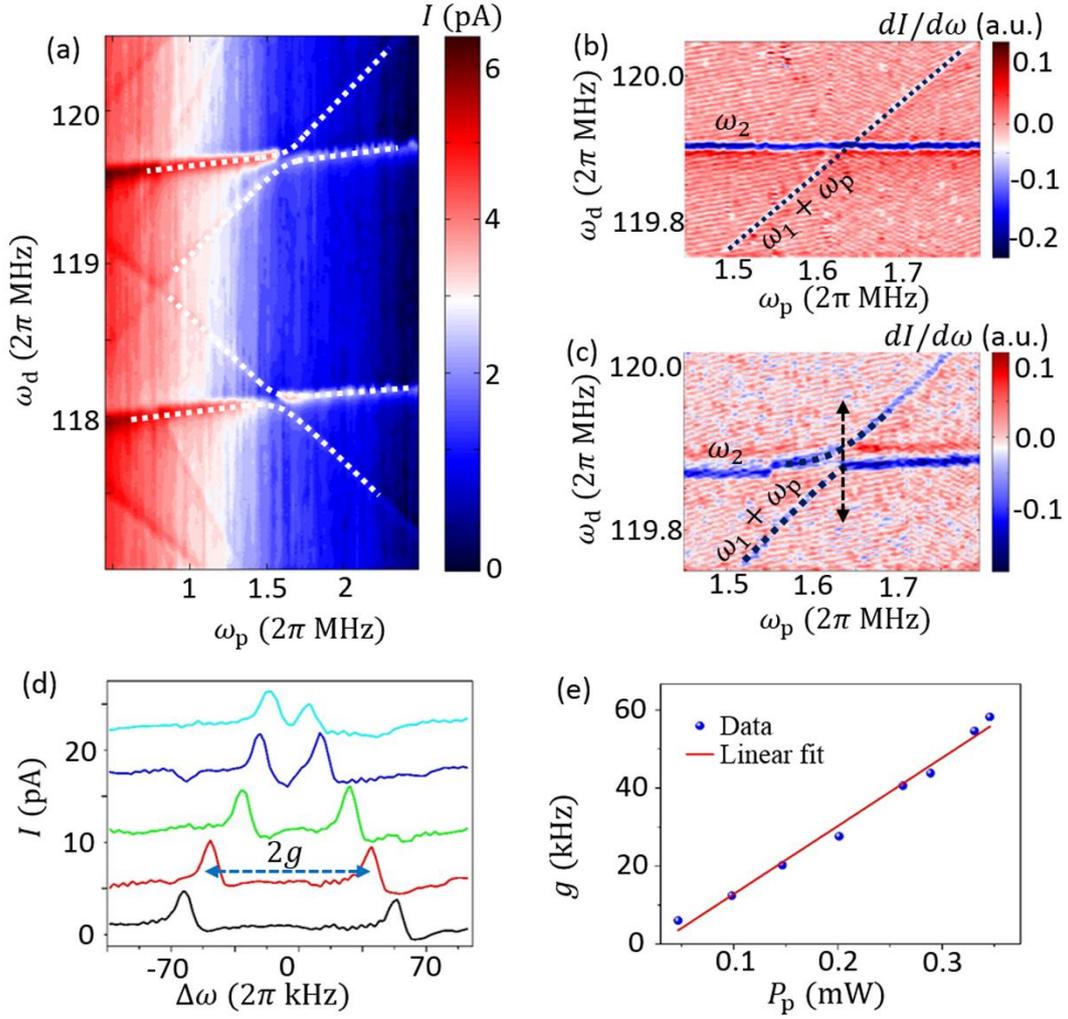

**Figure 4.** Tuning the two modes' coupling strength into the strong coupling regime. (a) Mode splitting with a pump power of $-10$ dBm. (b) Weak mode coupling with $-25$ dBm pump power, where a direct cross is observed. (c) Strong mode coupling with $-5$ dBm pump power, where a clear avoided mode splitting is observed. (d) Mode splitting for varioust pump powers. From bottom to top, the pump powers are: $0$, $-3$, $-6$, $-9$, and $-12$ dBm, each with 5 pA offset. (e) Coupling strength as a function of the pump power. The results are well fitted by a linear function when the pump power is in mW unit.

16